\documentclass[utf8]{FrontiersinHarvard} 

\usepackage{url,hyperref,microtype,subcaption}
\usepackage[onehalfspacing]{setspace}

\usepackage{tabularx}
\usepackage{multirow}
\usepackage{tablefootnote}

\def\keyFont{\fontsize{8}{11}\helveticabold }
\def\firstAuthorLast{Zhang {et~al.}} 
\def\Authors{Zhen Zhang\,$^{*}$, Yuantao Ding\,, Zhirong Huang\, and Feng Zhou}
\def\Address{SLAC National Accelerator Laboratory, Menlo Park, California 94025, USA }

\def\corrAuthor{Zhen Zhang}
\def\corrEmail{zzhang@slac.stanford.edu}

\begin{document}
\onecolumn
\firstpage{1}

\title {Multiplexed Photoinjector Optimization for High-Repetition-Rate Free-Electron Lasers} 

\author[\firstAuthorLast ]{\Authors} 
\address{} 
\correspondance{} 

\extraAuth{}

\maketitle

\begin{abstract}

\section{}
The multiplexing capabilities of superconducting X-ray free-electron lasers (FELs) have gained much attention in recent years. The demanding requirements for photon properties from multiple undulator lines necessitate more flexible beam manipulation techniques to achieve the goal of "beam on demand". In this paper, we investigate a multiplexed configuration for the photoinjector of high-repetition-rate FELs that aims to simultaneously provide low-emittance electron beams of different charges. A parallel, multi-objective genetic algorithm is implemented for the photoinjector parameter optimization. The proposed configuration could drastically enhance the flexibility of beam manipulation to improve multiplexing capabilities and realize the full potential of the facility.


\tiny
 \keyFont{ \section{Keywords:} free-electron laser, beam on demand, multiplexing capability, photoinjector, VHF gun, genetic algorithm} 
\end{abstract}

\section{Introduction}

Recently there is increasing interest in the multiplexing capabilities of high-repetition-rate X-ray free-electron lasers (FELs) across the community as the development of superconducting radio-frequency (SRF) linac based FEL facilities such as LCLS-II\,[\cite{raubenheimer2015lcls}] and its high energy upgrade\,[\cite{raubenheimer2018lcls}], SHINE\,[\cite{zhu2017sclf}] and European XFEL\,[\cite{decking2020mhz}]. The high-repetition rate of electron beams up to megahertz (MHz) in these facilities make it feasible to feed multiple undulator lines simultaneously, significantly increasing the user experiment time. The concept of generating bunches with different properties to drive more than one FELs in parallel has been already done both at superconducting FELs\,[\cite{faatz2016simultaneous,frohlich2019multi}] and normal conducting FELs\,[\cite{paraliev2022swissfel}].
However, the wide range of photon property requirements from multiple undulator lines presents a major challenge for satisfying these requirements with a single photoinjector and SRF linac.

To achieve the full potential of high-repetition-rate FEL facilities such as LCLS-II, the concept of ``beam on demand" has been proposed to provide tailored beam properties for each undulator line at the desired repetition rate\,\,[\cite{zhang2021beam}]. 
These properties include, but are not limited to, beam current, bunch length, beam charge, beam energy, and energy chirp.
To achieve shot-by-shot control of bunch length and peak current, a continuous-wave (CW) normal conducting cavity, called a ``chirper", which is located upstream of the first magnetic chicane, has been proposed and studied\,[\cite{nasr2016cw,zhang2023fast}].
 Different methods have been explored to produce the desired beam energy for each beamline from a single SRF linac, such as using an achromatic and isochronous electron delay system and fast kickers\,[\cite{yan2019multi}] or off-frequency detuning in SRF cavities\,[\cite{zhang2019multienergy}]. 
 Laser heater shaping allows for customized shaping of the beam's slice energy spread, resulting in various beam current profiles for different applications\,[\cite{marinelli2016optical,roussel2015multicolor,cesar2021electron}]. It's worth noting that all these techniques are performed after the photoinjector and assume consistent beam properties at the photoinjector exit for all shots.

The photoinjector is the low energy ($\lesssim$100\,MeV) part of the accelerator, where space charge effects and non-relativistic kinematics play an critical and even dominant role. The photoinjector is usually optimized for a specific beam charge, but changing the beam charge requires re-optimization of the entire injector settings. To achieve the goal of ``beam charge on demand", a multiplexed configuration for photoinjector is needed to preserve low emittance for different beam charges at the same time. Due to the intensity of space charge, most of the parameters in the photoinjector are nonlinearly coupled. In this paper, a simulation code, ASTRA, which has been extensively tested against experiments and other codes\,[\cite{flottmannAstra}], is used to model the beam dynamics in the photoinjector. A multi-objective genetic optimizer based on the continuous non-dominated sorting genetic algorithm (NSGA-II)\,[\cite{deb2002fast}] is used for the parameter optimizations of photoinjector. NSGA-II is a popular evolutionary algorithm that is widely used in various fields, including engineering, finance, and computer science. NSGA-II can handle multiple objectives and constraints simultaneously, and can effectively generate a set of Pareto-optimal solutions, which are the optimal solutions that cannot be improved in one objective without worsening another objective. The choice of population size in NSGA-II algorithm depends on several factors, including the complexity of the problem, the number of objectives, and the available computational resources. A larger population size can lead to better diversity in the population, which can improve the convergence and quality of the optimization results. However, a larger population size also requires more computational resources and longer computation time. This method has been widely adopted in the field of accelerator during the last decades\,[\cite{bazarov2005multivariate,hajima2006multiparameter,hofler2013innovative,papadopoulos2014rf2,chen2022beam,zhu2022inhibition,neveu2023comparison}]. In this study, the population size of each generation is set to be 128, taking into account the decision variables of photoinjector optimization and the available computational resources.

The paper is structured as follows: in Sec.~\ref{baseline}, we will present the optimized results of the LCLS-II photoinjector for three typical beam charges as a baseline. The method to optimize multiplexed configurations is described in Sec.~\ref{multilpexed}, including the optimal results of three multiplexed configurations with different customized knobs. Finally, in Sec.~\ref{conclusion}, we provide some discussion and concluding remarks.

\section{Baseline of LCLS-II photoinjector}\label{baseline}
\begin{figure}[h!]
\begin{center}
\includegraphics[width=18cm]{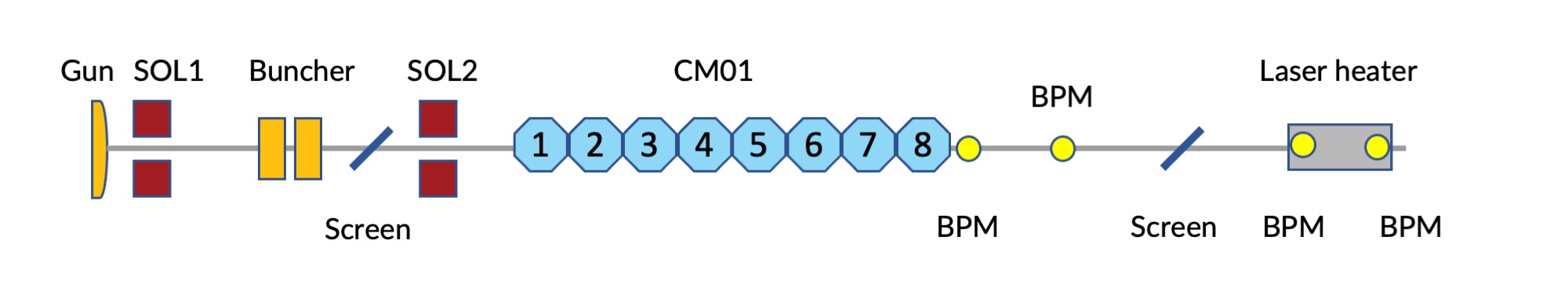}
\end{center}
\caption{  A schematic layout of the VHF gun photoinjector for LCLS-II.}\label{fig:layout}
\end{figure}

To illustrate the proposed multiplexed configuration for the photoinjector, we first show the optimization results for different beam charges individually optimized with the same input variables, serving as the facility's baseline. We use the photoinjector of LCLS-II as an example in the following simulations, which is depicted in Figure\,\ref{fig:layout}. The photoinjector consists primarily of a very-high-frequency (VHF) gun, two solenoids (labeled SOL1 and SOL2), a 2-cell normal conducting continuous-wave 1.3\,GHz buncher cavity, and eight 9-cell 1.3\,GHz cavities (labeled 1 to 8). The RF gun is a replica of the APEX\,[\cite{sannibale2012advanced}] gun developed at Lawrence Berkeley National Laboratory, which allows LCLS-II to operate in continuous-wave mode with an electron beam repetition rate of 1\,MHz. The drive laser used for generating electrons at the Cs$_2$Te cathode is an ultraviolet laser at 257.5\,nm, a fourth harmonic from an infrared laser operating at 1030\,nm\,[\cite{gilevich2020lcls}]. The longitudinal profile of the laser pulse can have either a Gaussian or uniform distribution. The implementation of the latter relies on successful laser pulse shaping to suppress any initial density modulation, as failure to do so may introduce a strong microbunching instability in the accelerator\,[\cite{huang2004suppression}] In this study, we assume a uniform transverse and longitudinal laser density profile. The laser spot radius and pulse duration are also considered as optimization variables. The maximum gradient of the gun is fixed at 20\,MV/m in the simulations, equivalent to a maximum beam energy of 750\,keV. The gradient of the buncher cavity is also fixed to produce a maximum energy gain of 200\,keV. Both the gun and buncher's relative acceleration phases are optimization variables. The two solenoids are utilized to control the beam size along the photoinjector and implement the emittance compensation technique\,[\cite{carlsten1989new,serafini1997envelope}]. For the downstream cryomodule, the first three cavities are considered as optimization variables, while the others are set to be on-crest acceleration with a fixed gradient.

A high brightness electron beam at the exit of the photoinjector is a crucial requirement. This is typically measured by the transverse emittance and the bunch length, which have an inverse relationship. Minimizing both of these parameters in the optimization process typically results in a ``Pareto front," which represents the minimum possible transverse emittance for a given bunch length.
Aside from these two objectives, other constraints must also be considered to ensure that the results are feasible. These include the total beam energy and projected energy spread, which must be within acceptable limits for the downstream laser heater. For high-repetition rate FELs, it is also important to consider higher-order ($>$2) correlated energy spread, as it can lead to current horns after beam compression. In this study, we focus on optimizing the transverse emittance and bunch length as our two main objectives. We maintain cylindrical symmetry throughout the photoinjector, so only the horizontal emittance ($\epsilon_{nx}$) was used as our metric for transverse emittance. Other constraints will be examined to ensure they fall within reasonable limits.

In our simulations, the mean transverse energy (MTE) of electrons at cathode emission is 330\,meV, which corresponds to an initial intrinsic emittance of 0.8 mm-mrad/mm (also known as thermal emittance). This estimate of the intrinsic emittance is considered to be conservative and reasonable for Cs$_2$Te cathodes. To balance the computational resources and accuracy, we use 10,000 macro particles in each simulation run with ASTRA, and a grid of 25 by 50 in the radial and longitudinal directions, respectively. Once the optimization has converged, we select one of the optimized solutions from the Pareto front and run it again with higher accuracy, using 200,000 particles and a grid of 75 by 125 to minimize numerical errors. Notably, we have observed good agreement between simulations with both low and high numbers of particles, with differences being negligible for our study.

The variables and constants used in the simulation of the LCLS-II photoinjector are listed in Table,\ref{table:pars}. The drive laser pulse is assumed to have a uniform radial shape and density profile in the transverse direction. Previous optimization work has shown that the gradient of the second cavity (CAV-2) in the cryomodule is very low ($<$5\,MV/m) for emittance compensation [\cite{mitchell2016rf},\cite{neveu2020lcls}], and is purposely turned off during LCLS-II commissioning. For simplicity, we also assume on-crest acceleration with fixed gradient from CAV-4 to CAV-8, as these cavities have little effect on the beam emittance and bunch length. During the machine operation, the phases and amplitudes of these cavities can be further adjusted to correct the beam energy and correlated energy spread based on the measurements.

\begin{figure}[h!]
\begin{center}
\includegraphics[width=14cm]{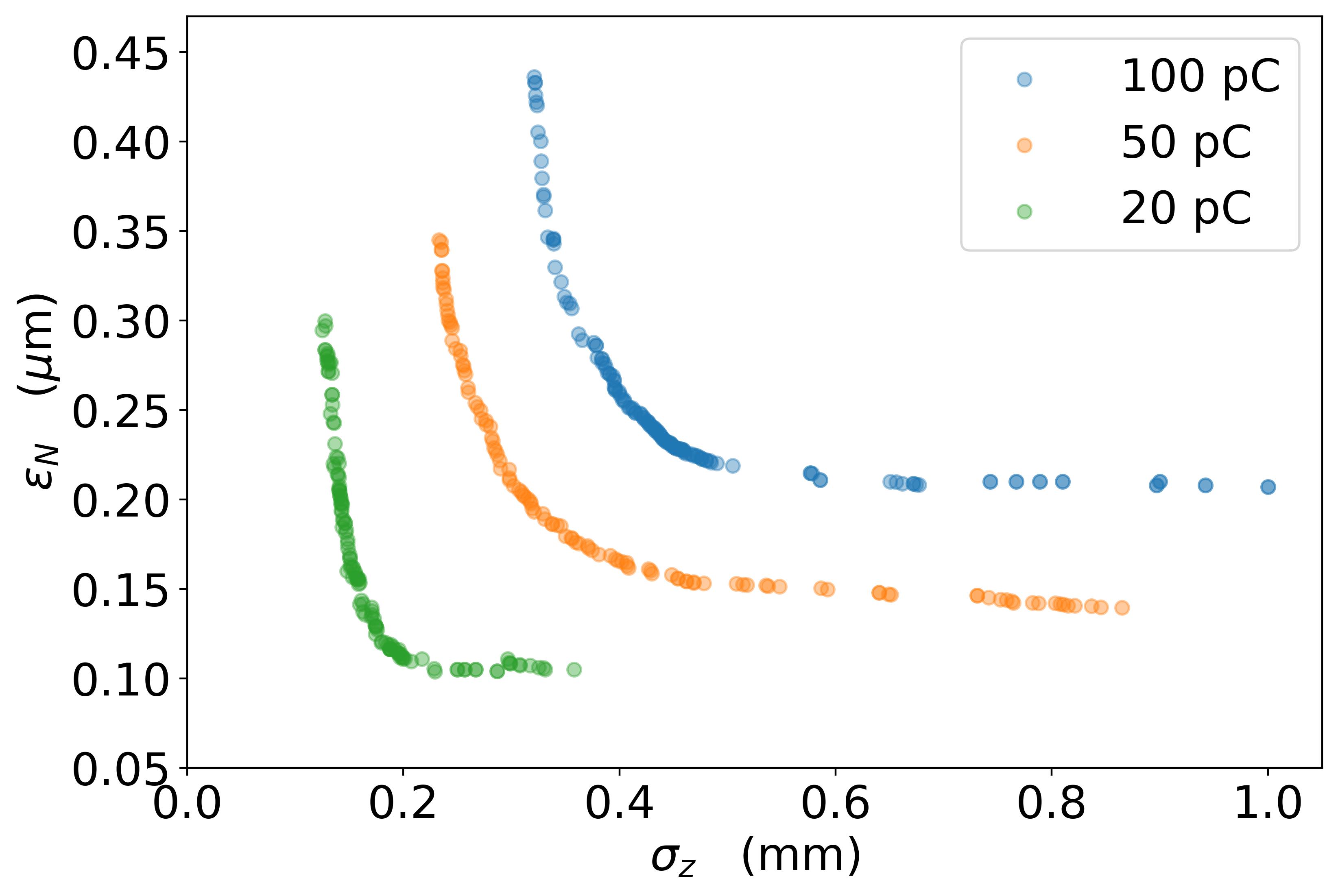}
\end{center}
\caption{ Pareto fronts of transverse emittance $\epsilon_{nx}$ and rms bunch length $\sigma_z$ for individual configurations with different beam charges.}\label{fig:pareto:individual}
\end{figure}

In this study, we explore the performance of the LCLS-II photoinjector for three different beam charges: 100\,pC, 50\,pC, and\,20 pC. To find the optimal combination of transverse emittance and bunch length, we conduct two-objective optimizations for each of these beam charges. The results are displayed in Figure,\ref{fig:pareto:individual}, which shows the Pareto fronts of the two objectives for each beam charge. The Pareto fronts are obtained after 150 generations with 128 samples per generation. As a result of the optimizations, the following values of transverse emittance and bunch length are selected as baseline solutions: 0.22 $\mu$m at $\sigma_z$ = 0.8 mm for 100 pC, 0.15 $\mu$m at $\sigma_z$ = 0.5 mm for 50 pC, and 0.10 $\mu$m at $\sigma_z$ = 0.3 mm for 20 pC. These solutions will be used as the reference points for following optimizations of the multiplexed configuration. We note that while the Pareto fronts in Figure \ref{fig:pareto:individual} suggest that shorter bunch lengths can be achieved with a slight increase in emittance, shorter bunch lengths can also lead to stronger high-order energy chirp on the beam due to space charge force, which may ultimately degrade beam properties after downstream compression. As such, we prioritize selecting a reasonable range for the bunch length, rather than choosing the shortest possible value.

\section{Multiplexed configuration for LCLS-II photoinjector}\label{multilpexed}
The objective of the multiplexed configuration for photoinjector is to produce high-brightness electron beams with different beam charges (100\,pC, 50\,pC and 20\,pC) alternatively at high repetition rate. To achieve this goal, certain settings in the photoinjector must remain constant for all beam charges, such as the field of the solenoid and the gradient of the RF cavities. The most challenging task in this work is to preserve low emittance for all charges with limited knobs. To simplify the optimization and reduce the number of objectives, the bunch lengths of different charges are specified at specific values and the emittance is minimized. A loss function $L_q$ is defined for a given beam charge $q$ as
\begin{align}
L_q = \epsilon_{nx} + max(abs(\sigma_z - \sigma_z^{(q)})-r_q,0)\times\lambda\,,
\end{align}
where $\epsilon_{nx}$ is the value of emittance in unit of $\mu$m and $\sigma_z$ is the rms bunch length in unit of mm. When the bunch length is within the range of $\sigma_z^{(q)}\pm r_q$, the loss function is exactly the value of emittance. $\lambda$ is a regularization parameter to control the penalty when bunch length is out of the desired range. In our simulations, we choose $\lambda=2$. These loss functions of different beam charges are used as objectives in optimizations. In this work, we consider two groups of target bunch lengths: (I) 1.0$\pm$0.1\,mm for 100\,pC, 0.8$\pm$0.1\,mm for 50\,pC, 0.6$\pm$0.1\,mm for 20\,pC; (II) 0.8$\pm$0.1\,mm for 100\,pC, 0.5$\pm$0.075\,mm for 50\,pC, 0.3$\pm$0.05\,mm for 20\,pC.

In the multiplexed configuration of the photoinjector, several parameters associated with magnetic and RF fields must remain constant for all beam charges. However, other parameters, primarily those related to the drive laser pulse, can be adjusted as needed for different charges. These adjustable parameters include the laser spot size at the cathode, laser pulse duration, and laser injection time. Our preliminary studies have shown that the laser injection time is not a significant knob in preserving the emittance. This can be attributed to the fact that the wavelength of the VHF gun is much longer than that of the downstream RF cavities and the laser pulse duration, and thus, adjusting the laser injection time within a reasonable range for the downstream RF cavities does not significantly affect the space charge effect in the low-energy electron beam, i.e., the emittance compensation process. In this work, we consider three cases: (A) no customization of parameters, i.e., all variables in Table~.\ref{table:pars} are adjusted together and kept the same for the 3 bunch charges; (B) customization of the laser spot size, allowing the laser spot size as a free variable for each charge while keeping all other variables the same for the three charges; and (C) customization of laser pulse duration, which is similar with (B) but allowing the laser pulse duration as a free variable for each charge. The optimal solutions will be compared with the results obtained in the previous section for individual configurations.

\begin{figure}[h!]
\begin{center}
\includegraphics[width=14cm]{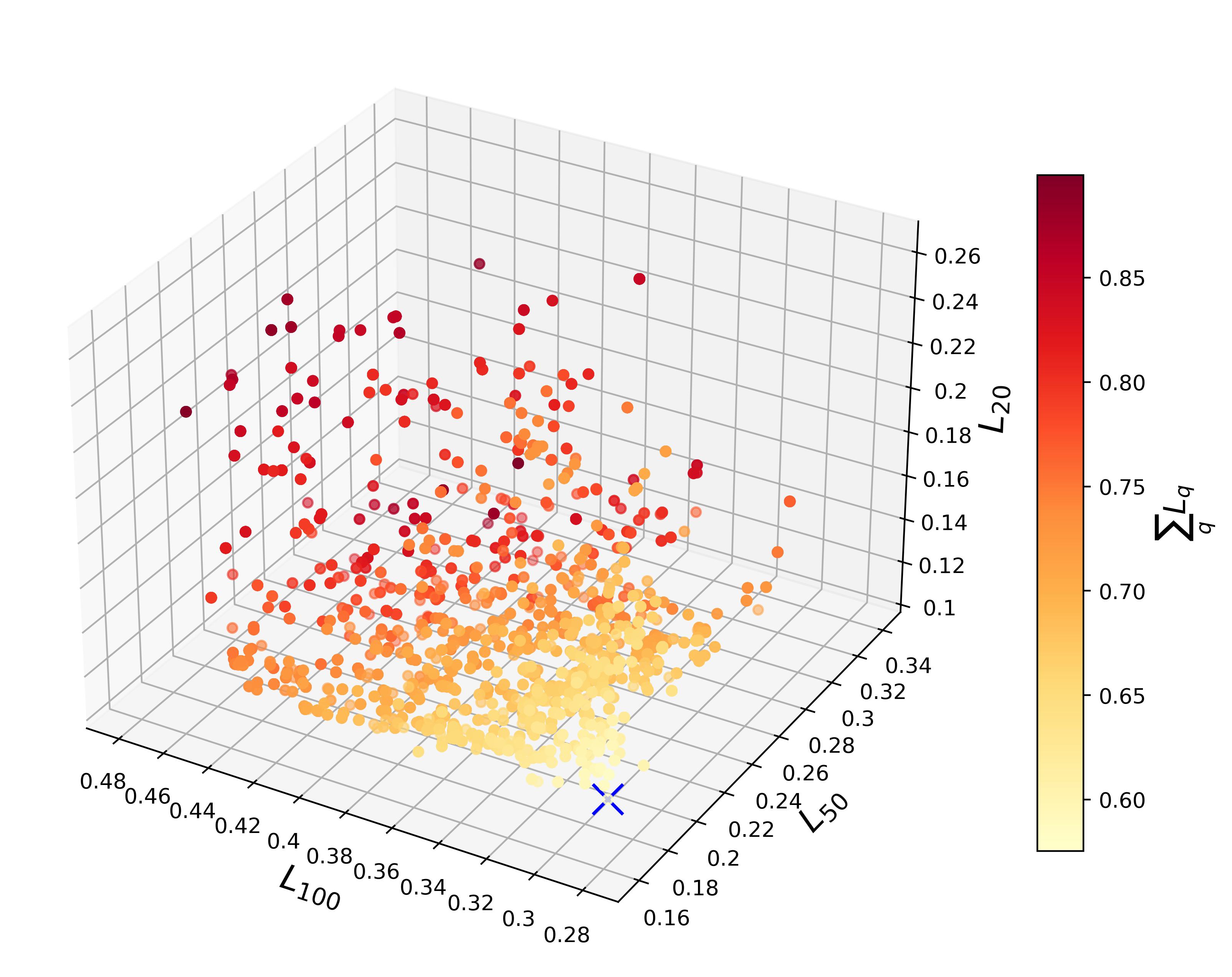}
\end{center}
\caption{ Loss function $L_q$ distribution of three beam charges of the last 40 generations in the optimization of multiplexed configuration with customized laser spot size. The color map means the sum of three loss functions. The blue cross represents the optimal point with minimum sum of loss functions.}\label{fig:pareto:3d}
\end{figure}

Figure\,\ref{fig:pareto:3d} displays the distribution of the three loss functions during the optimization of the multiplexed configuration with a customized laser spot size. The color map represents the sum of the three loss functions. The optimization process converges after approximately 200 generations, and the optimal solution, with the minimum sum of the three loss functions, is indicated in the figure by a blue cross.

\begin{figure}[h!]
\begin{center}
\includegraphics[width=14cm]{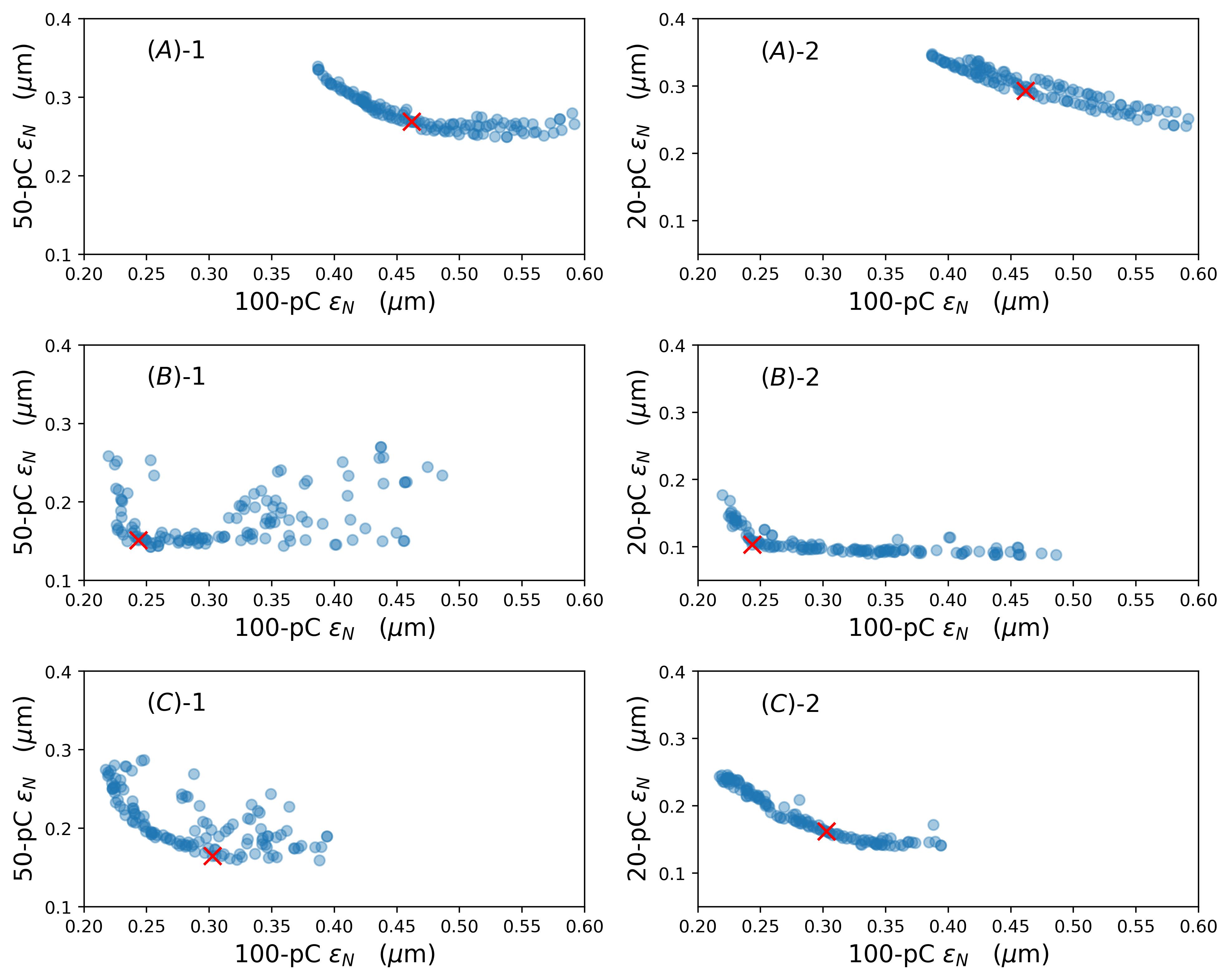}
\end{center}
\caption{ Distribution of beam emittance between 100\,pC and 50\,pC (left column), 100\,pC and 20\,pC (right column) in the multiplexed configuration. The desired rms bunch length is 1.0$\pm$0.1\,mm for 100\,pC, 0.8$\pm$0.1\,mm for 50\,pC and 0.6$\pm$0.1\,mm for 20\,pC, respectively, in the loss functions. (Top) case A: no customized knob. (Middle) case B: customized laser spot size. (Bottom): case C: customized laser pulse duration. The red cross represents the optimal point with minimum sum of the emittance of three beam charges in each configuration.}\label{fig:3_charge}
\end{figure}

For a given desired bunch length range (e.g., group I), the emittance distribution for each of the two beam charges for the three cases is shown in Figure\,\ref{fig:3_charge}. A clear "Pareto front" between the emittances of each pair of beam charges can be observed. In each configuration, the optimal solution is chosen to minimize the sum of emittances, as indicated by a red cross in the figure. The detailed bunch lengths and emittances of these three configurations are listed in Table\,\ref{table:results}. Both configurations with a customized knob perform better than the one without any knobs. The multiplexed configuration with a customized laser spot size (case B) outperforms the other two configurations, with a total increase of around 9\% in the sum of emittances, compared to the baseline individually optimized configurations for each beam charge. The emittances of the 50\,pC and 20\,pC beams remained almost the same as the baseline. The performance of the configuration with a customized laser pulse duration (case C) is not as good as that of case B, but it is still much better than the one without any customizations (case A). For different bunch length combinations, the total emittance increase of the multiplexed configuration may vary, but it is crucial that they remain within a reasonable range for facility operation.

From these three cases, we can see that a solution of a unified injector configuration working for all the three bunch charges can be found. If the laser parameters can be varied, the laser spot size is the most effective knob for optimizing the multiplexed configuration. Among the two desired bunch length ranges, the multiplexed configuration performs better for group I, which suggests that choosing an appropriate desired bunch length range can further reduce the increase in emittance in the multiplexed configuration.

In the multiplexed configuration with customized laser spot size (Figure\,\ref{fig:3_charge}-B), the optimal laser spot radius for 100\,pC, 50\,pC and 20\,pC are found to be 0.29\,mm, 0.21\,mm, and 0.12\,mm, respectively. The results show that the ratio of the square of the laser spot radius to the beam charge is nearly constant, meaning that the laser spot size is adjusted to keep the same charge density near the cathode emission, thus preserving emittance. This is a significant finding as it implies that the multiplexed configuration can be easily adapted to other beam charges beyond the three beam charges used in the optimization study.

\begin{figure}[h!]
\begin{center}
\includegraphics[width=14cm]{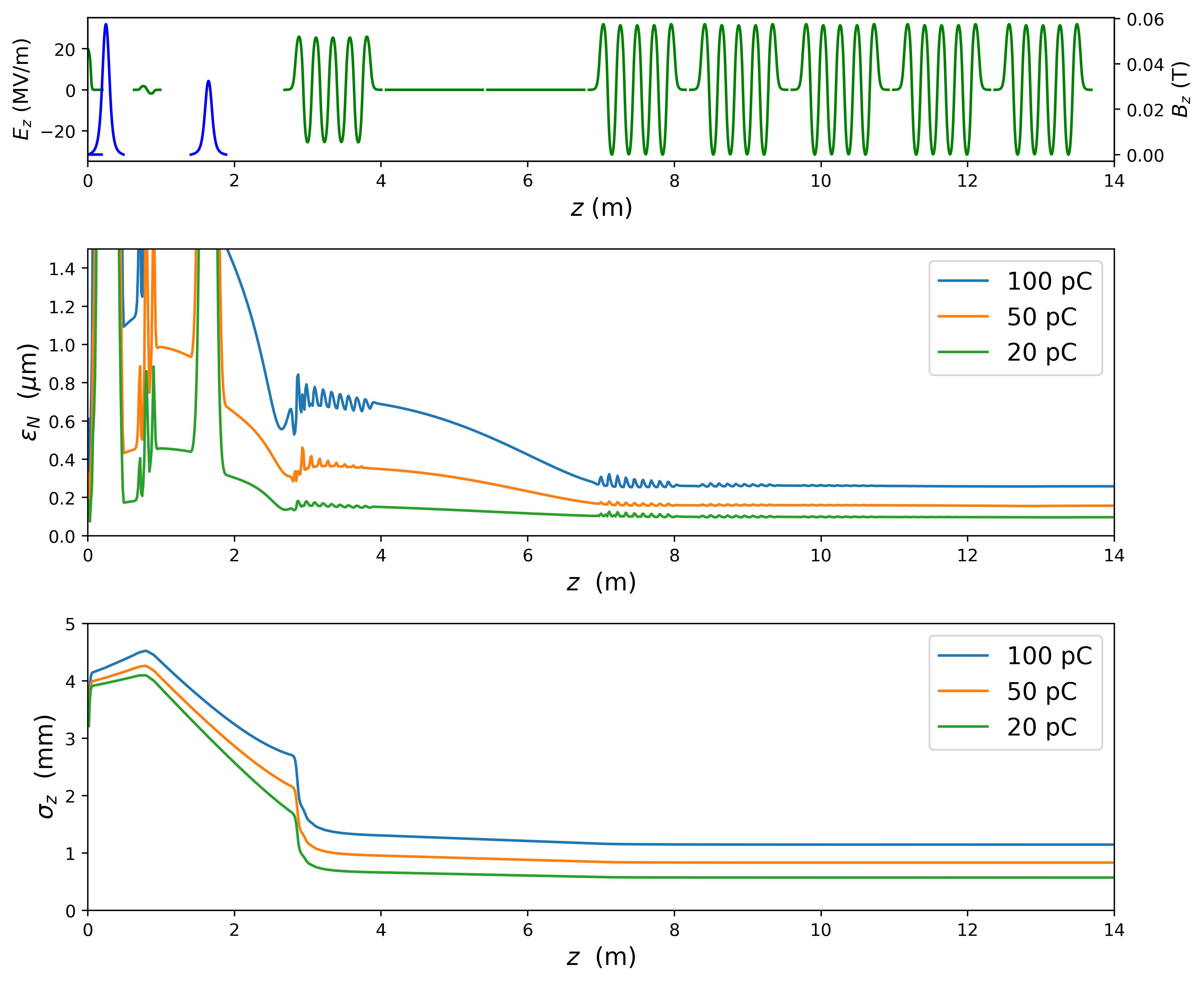}
\end{center}
\caption{ Electric and solenoid field (top), normalized emittance (middle) and rms bunch length (bottom) of three beam charges in the multiplexed configuration with customized laser spot size (case B in group I).}\label{fig:field_emit_sigz}
\end{figure}

\begin{figure}[h!]
\begin{center}
\includegraphics[width=14cm]{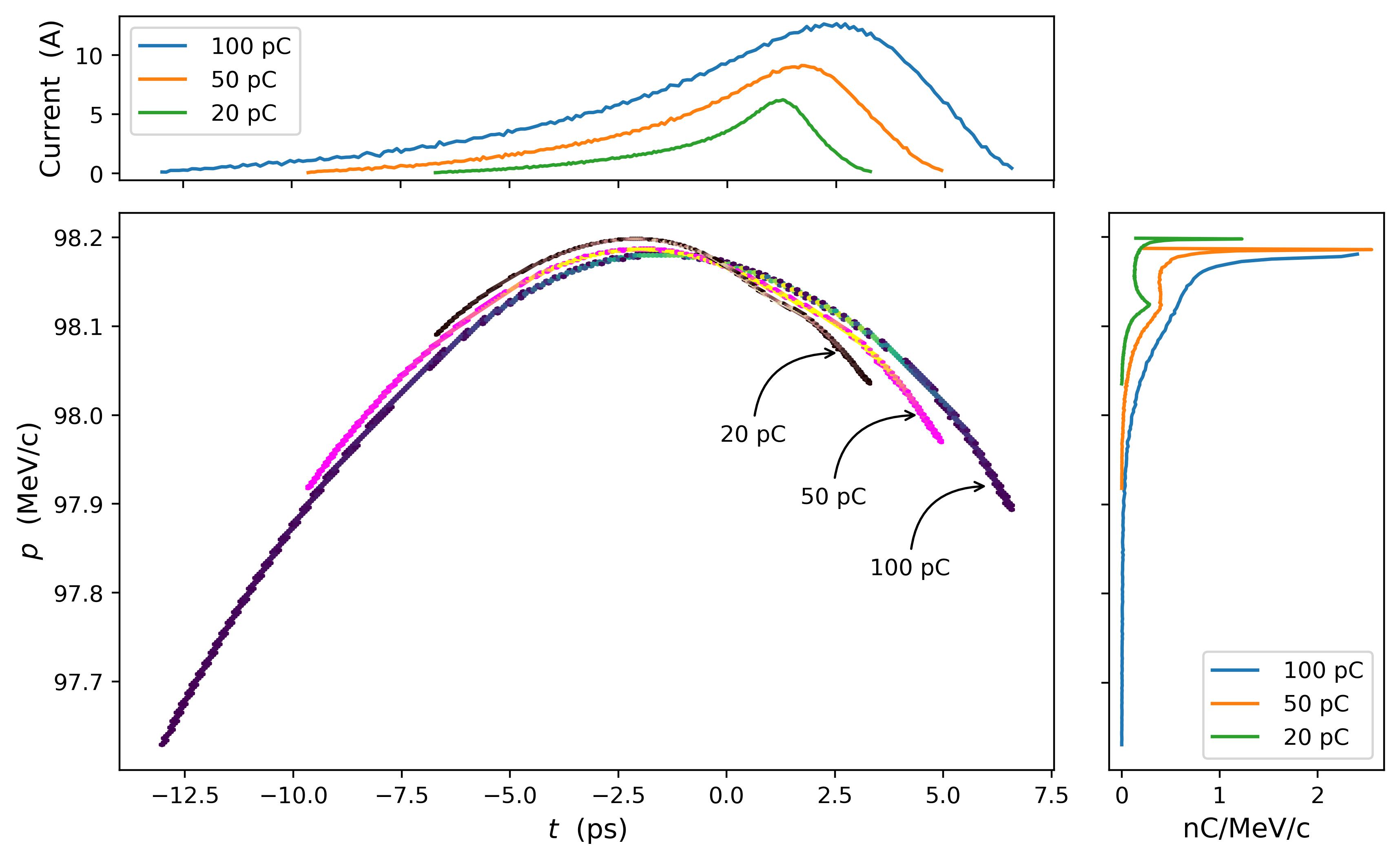}
\end{center}
\caption{ Longitudinal phase space of electron beams at the exit of photoinjector of three beam charges in the multiplexed configuration with customized laser spot size (case B in group I). To distinguish between each beam, we used a different color map for density representation, which is not shown in the figure.}\label{fig:lps}
\end{figure}

For the optimal settings of case B in group I, the evolution of the emittance and the rms bunch length, as well as the electric and solenoid fields, are displayed in Figure\,\ref{fig:field_emit_sigz}. The emittance reaches its minimum value just before being frozen at high beam energy, and the rms bunch length stays constant after CAV-1. Although there is some correlated energy chirp on the beam at the end of the injector if all the last 5 cavities are set at on-crest acceleration, it can be removed by shifting their acceleration phase by approximately 6\,degrees. The longitudinal phase space of three beams and their current and energy profiles are shown in Figure\,\ref{fig:lps}. It can be observed that the average beam energy and arrival time of these three beams are very close, a result of the same acceleration phases of all RF cavities and the same laser pulse duration. Different beam charges may result in different wakefields along the SRF linac and therefore different beam compression in the downstream chicane compressors, but this can be adjusted shot-by-shot using the proposed chirper cavity for LCLS-II and its high energy upgrade\,[\cite{zhang2023fast}].

\section{Discussions and conclusions}\label{conclusion}
In this study, we investigated the possibility of multiplexed configurations for the photoinjector of SRF-based FELs to deliver low-emittance electron beams with different beam charges at high repetition rates. We evaluated three different configurations, two of them with a different customization of the drive laser system for different beam charges. Our results showed that the laser spot size, which can be technically customizable for each beam charge, was the most effective factor in preserving beam emittance. Additionally, we found that maintaining the same charge density on the cathode by adjusting the laser spot size made it possible to apply the optimized multiplexed configuration to other beam charges. The implementation of these multiplexed configurations in the photoinjector of SRF-based FELs will enable the delivery of beam charges on demand for each individual beamline, thus maximizing the multiplexing capabilities of the facilities.


\section*{Author Contributions}

Z.Z., Y.D., Z.H. and F.Z. contributed to conception and design of the study. Z.Z. performed the simulations and the statistical analysis. Z.Z. and Y.D. wrote the first draft of the manuscript. Z.Z., Y.D., Z.H. and F.Z. contributed to the manuscript revision. All authors approved the publication of the content. 

\section*{Funding}
This work was supported by U.S. Department of Energy Contract No. DE-AC02-76SF00515.

\section*{Acknowledgments}
The authors would like to thank Jingyi Tang and Christopher Mayes from SLAC for their valuable help and discussions on the optimizations of photoinjector with genetic algorithm. 

\section*{Data Availability Statement}
Data will be made available on request.

\bibliographystyle{Frontiers-Harvard} 
\bibliography{test}


\section*{Figure captions}


\begin{table}[h!]
  \caption{Optimization variables and constants in the LCLS-II photoinjector.}
  \label{table:pars}
  \centering
  \begin{tabularx}{.8\textwidth}{p{0.4\textwidth}p{0.25\textwidth}p{0.15\textwidth}}
    \hline\hline
    \textbf{Variable / Constant} & \textbf{Value} & \textbf{Unit} \\
    \hline
    Laser transverse distribution & radial uniform & / \\
    \hline
    Laser spot radius  & 0.1 $\sim$ 1 & mm \\
    \hline
    Laser longitudinal distribution & uniform & / \\
    \hline
    Laser pulse duration (rms$^1$) & 5 $\sim$ 24 & ps \\
    \hline
    Beam charge & 100 / 50 / 20 & pC \\
    \hline
    VHF gun gradient & 20 & MV/m \\
    \hline
    VHF gun off-crest phase & -30 $\sim$ 30 & degree $^2$\\
    \hline
    Buncher cavity gradient & 1.8 & MV/m \\
    \hline
    Buncher cavity off-crest phase & -120 $\sim$ -40 & degree \\
    \hline
    CAV-1 and CAV-3 peak gradient$^3$ & 10 $\sim$ 32 or turn off & MV/m \\
    \hline
    CAV-1 and CAV-3 off-crest phase & -60 $\sim$ 40 & degree \\
    \hline
    CAV-2 & Turn off & / \\
    \hline
    CAV-4 to CAV-8 peak gradient & 32 & MV/m \\
    \hline
    CAV-4 to CAV-8 off-crest phase & 0 & degree \\
    \hline
    SOL1 B peak field & 0.05 $\sim$ 0.065 & T \\
    \hline
    SOL2 B peak field & 0.02 $\sim$ 0.04 & T \\
    \hline\hline
    \multicolumn{3}{X}{\footnotesize{$^1$ RMS width equals full width divided by $\sqrt{12}$ in uniform distribution.}}\\
    \multicolumn{3}{X}{\footnotesize{$^2$ All phases are based on the corresponding RF wavelength.}} \\
    \multicolumn{3}{X}{\footnotesize{$^3$ Average gradient of CM cavity equals peak gradient divided by 1.93.}} \\
  \end{tabularx}
\end{table}

\begin{table}[h!]
  \caption{ Electron bunch length and the corresponding optimal emittance of three beam charges in the baseline and three multiplexed configurations. Case (A) no customization of parameters for the 3 bunch charges; case (B) customization of the laser spot size for each charge; and case (C) customization of laser pulse duration for each charge. Details see the text.}
  \label{table:results}
  \centering
  \begin{tabularx}{.88\textwidth}{p{0.11\textwidth}|p{0.11\textwidth}|p{0.06\textwidth}|p{0.11\textwidth}|p{0.11\textwidth}|p{0.11\textwidth}|p{0.11\textwidth}}
    \hline\hline
    \textbf{Group No.} & \textbf{Parameter} & \textbf{Unit} & \textbf{Baseline} & \textbf{Case A}  & \textbf{Case B} & \textbf{Case C}\\
    \hline
    \multirow{7}{3em}{I} & 100-pC $\sigma_z$ & mm & 1.00 & 1.08 & 1.10 & 1.06\\
     & 50-pC $\sigma_z$ & mm & 0.80 & 0.87 & 0.80 & 0.83\\
     & 20-pC $\sigma_z$ & mm & 0.60 & 0.58 & 0.55 & 0.69 \\
     \cline{2-7}
     & 100-pC $\epsilon_n$ & $\mu$m & 0.20 & 0.46 & 0.24 & 0.30 \\
     & 50-pC $\epsilon_n$ & $\mu$m & 0.15 & 0.27 & 0.15 & 0.16 \\
     & 20-pC $\epsilon_n$ & $\mu$m & 0.10 & 0.29 & 0.10 & 0.16 \\
     & sum of $\epsilon_n$ & $\mu$m & 0.45 & 1.02 & 0.49 & 0.62 \\
    \hline
    \multirow{7}{3em}{II} & 100-pC $\sigma_z$ & mm & 0.80 & 0.83 & 0.86 & 0.72\\
     & 50-pC $\sigma_z$ & mm & 0.50 & 0.55 & 0.58 & 0.55\\
     & 20-pC $\sigma_z$ & mm & 0.30 & 0.29 & 0.33 & 0.35 \\
     \cline{2-7}
     & 100-pC $\epsilon_n$ & $\mu$m & 0.21 & 0.33 & 0.28 & 0.36 \\
     & 50-pC $\epsilon_n$ & $\mu$m & 0.15 & 0.38 & 0.17 & 0.20 \\
     & 20-pC $\epsilon_n$ & $\mu$m & 0.10 & 0.33 & 0.12 & 0.18 \\
     & sum of $\epsilon_n$ & $\mu$m & 0.46 & 1.04 & 0.57 & 0.74 \\
     \hline\hline
  \end{tabularx}
\end{table}

\end{document}